\documentclass[useAMS,usenatbib,A4]{mn2e}

\usepackage{graphicx}  
\usepackage{ amssymb }
\usepackage{journals}
\usepackage[dvips]{color}
\newcommand{\kms}{km s$^{-1}$}

\newcommand{\feii}{Fe\,{\sc ii}}

\newcommand{\caii}{Ca\,{\sc ii}}

\newcommand\simgt{\lower.3ex\hbox{\gtsima}}

\usepackage{placeins}
\usepackage[modulo]{lineno}

\title[The abundance spread in NGC 6273]{Confirming the intrinsic abundance
spread in the globular cluster NGC 6273 (M 19) with calcium triplet
spectroscopy\thanks{Based on observations obtained at the Gemini Observatory,
which is operated by the Association of Universities for Research in Astronomy,
Inc., under a cooperative agreement with the NSF on behalf of the Gemini
partnership: the National Science Foundation (United States), the National
Research Council (Canada), CONICYT (Chile), the Australian Research Council
(Australia), Minist\'{e}rio da Ci\^{e}ncia, Tecnologia e Inova\c{c}\~{a}o
(Brazil) and Ministerio de Ciencia, Tecnolog\'{i}a e Innovaci\'{o}n Productiva
(Argentina).}} \author[D.\ Yong, G.\ S.\ Da Costa and J.\ E.\ Norris] 
{David Yong,$^{1}$\thanks{E-mail: david.yong@anu.edu.au}
Gary S.\ Da Costa$^{1}$ and 
John E.\ Norris$^{1}$. \\
\\ 
$^{1}$Research School of Astronomy and Astrophysics, Australian
National University, Canberra, ACT 2611, Australia 
}
\begin{document}


\pagerange{\pageref{firstpage}$-$\pageref{lastpage}} \pubyear{2015}

\maketitle

\label{firstpage}

\begin{abstract}

We present metallicities for red giant stars in the globular cluster NGC 6273
based on intermediate resolution GMOS-S spectra of the calcium triplet region.
For the 42 radial velocity members with reliable calcium triplet line strength
measurements, we obtain metallicities, [Fe/H], using calibrations established
from standard globular clusters. We confirm the presence of an intrinsic
abundance dispersion identified by \citet{Johnson:2015aa}. The total range in
[Fe/H] is $\sim$1.0 dex and after taking into account the measurement errors,
the intrinsic abundance dispersion is $\sigma_{\rm int}$[Fe/H] = 0.17 dex.
Among the Galactic globular clusters, the abundance dispersion in NGC 6273 is
only exceeded by $\omega$ Cen, which is regarded as the remnant of a disrupted
dwarf galaxy, and M 54, which is the nuclear star cluster of the Sagittarius
dwarf galaxy. If these three globular clusters share the same formation
mechanism, then NGC 6273 may also be the remnant nucleus of a disrupted dwarf
galaxy. 

\end{abstract}

\begin{keywords}
stars: abundances – stars: Population II – globular clusters: general – globular
clusters: individual: NGC 6273.
\end{keywords}

\section{INTRODUCTION}

The Milky Way Galaxy's most massive globular cluster, $\omega$ Centauri,
harbours a large star-to-star variation in metallicity, [Fe/H]\footnote{We use
the standard spectroscopic notation; [Fe/H] = $\log_{10}(\rm N_{\rm Fe}/N_{\rm
H})_\star - \log_{10}(N_{\rm Fe}/N_{\rm H})_\odot$}, spanning at least a factor
of 10 (e.g., \citealt{Freeman:1975aa,Norris:1995aa}). For the light, $\alpha$-,
Fe-peak and slow neutron-capture process ($s$-process) elements, $\omega$ Cen
also exhibits a complex distribution in abundance ratios and [X/Fe] (e.g.,
\citealt{Smith:2000aa,Cunha:2002aa,Stanford:2007aa,Johnson:2010aa,DOrazi:2011aa,Marino:2011aa,Pancino:2011aa}).
That is, $\omega$ Cen has experienced a complex chemical enrichment history and
has retained ejecta from asymptotic giant branch (AGB) stars as well as from
Type Ia and Type II supernovae \citep{Romano:2007aa,Marcolini:2007aa}. In
contrast, the majority of Galactic globular clusters do not exhibit
star-to-star variations in metallicity beyond the measurement uncertainties
(e.g., \citealt{Carretta:2009ac}). At present, arguably the most plausible 
explanation for $\omega$ Cen is that it is the surviving nucleus of an accreted
dwarf galaxy \citep{Freeman:1993aa,Lee:1999aa,Bekki:2003aa}. 

As recently as 2007, $\omega$ Centauri was the only globular cluster for which
there was undisputed evidence for a metallicity variation. Within the past few
years, however, abundance dispersions in Fe-peak and/or $s$-process elements
have been reported in a number of globular clusters including NGC 1851
\citep{Yong:2008ab,Villanova:2010aa,Carretta:2011aa,Gratton:2012ac,Marino:2014aa},
M 22 \citep{Marino:2009aa,Marino:2011ab,Da-Costa:2009aa,Roederer:2011aa},
Terzan 5 \citep{Ferraro:2009aa,Origlia:2011aa,Origlia:2013aa}, M 54
\citep{Carretta:2010ab,Carretta:2010ac},  M 75 \citep{Kacharov:2013aa}, NGC
5824 \citep{Da-Costa:2014aa}, M 2 \citep{Yong:2014aa} and NGC 5286
\citep{Marino:2015aa}. (Results for some of these clusters have been challenged
by, e.g.,  \citealt{Mucciarelli:2015aa} and \citealt{Roederer:2016aa}.)  

It was recognised that these objects (along with $\omega$ Cen), are
preferentially among the most luminous (i.e., massive) of the Galaxy's globular
clusters, and that many have an extended horizontal branch with extremely blue
stars \citep{Lee:2007aa}. Some, if not all, of these objects could be the
remnants of dwarf galaxies \citep{Bekki:2012aa,Da-Costa:2015aa,Marino:2015aa}.
In particular, M 54 is currently the nuclear star cluster of the Sagittarius
(Sgr) dwarf spheroidal galaxy (dSph). \citet{Carretta:2010ac} proposed that
when the Sgr dSph is tidally disrupted, the compact remnant of the M 54+Sgr
system will resemble $\omega$ Cen. Therefore, the identification of additional
globular clusters with intrinsic abundance dispersions in Fe-peak elements may
have important consequences for our understanding of Galactic evolution. In
order to complete the characterisation of the Galactic globular cluster system
and possibly identify new clusters with abundance dispersions in Fe-peak
elements, the most promising candidates to investigate are the most luminous
globular clusters with extended blue horizontal branches. 

NGC 6273 (M 19) has an extended blue horizontal branch and with $M_V$ =
$-$9.1, it is the 10th most luminous Galactic globular cluster \citep[][updated
December 2010]{Harris:1996aa} and lies near the Galactic bulge. Proper-motion
analysis indicates that it is an inner halo globular cluster
\citep{Casetti-Dinescu:2010aa}. Observations by \citet[][hereafter
J15]{Johnson:2015aa} revealed an intrinsic variation in Fe-peak and $s$-process
element abundances.  J15 found a ``metal-poor'' population (9 stars,
$\langle$[Fe/H]$\rangle$ = $-$1.75), a ``metal-rich'' population (8 stars,
$\langle$[Fe/H]$\rangle$ = $-$1.51) and one ``anomalous'' star with [Fe/H] =
$-$1.30. The ``metal-rich'' population had enhanced ratios of the $s$-process
element La compared to the ``metal-poor'' population, $\langle$[La/Fe]$\rangle$
= +0.16 and $\langle$[La/Fe]$\rangle$ = +0.48, respectively. Therefore, NGC
6273 has experienced a complex chemical enrichment history with contributions
from Type II supernovae and AGB stars. Narrowband Ca photometry by
\citet{Han:2015aa} supports the results from J15. The goal of this work is to
study a large sample of red giant branch (RGB) stars in NGC 6273 in order to
confirm and quantify the abundance dispersion. In Section 2 we describe the
sample selection, observations, data reduction and analysis. The results are
presented in Section 3. Sections 4 and 5 include our discussion and
conclusions. 

\section{SAMPLE SELECTION, OBSERVATIONS, REDUCTION AND ANALYSIS}

\begin{figure}
\centering
      \includegraphics[width=.99\hsize]{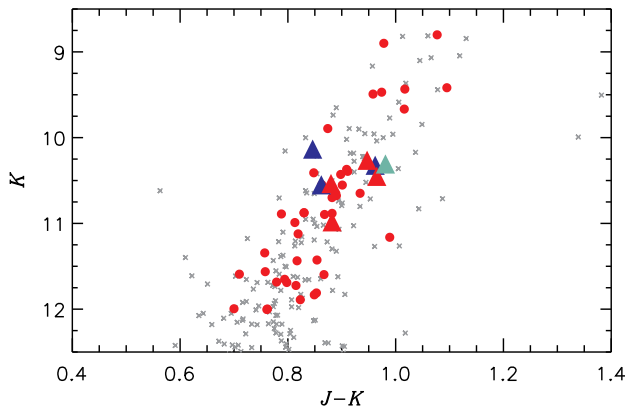} 
      \caption{$J-K$ colour-magnitude diagram for NGC 6273 based on 2MASS
photometry \citep{Skrutskie:2006aa}. All objects lie within a radius of 3
arcmin and have ‘AAA’ photometric quality flags. Red circles are likely radial
velocity members with reliable \caii\ triplet EW measurements. The blue, red
and aqua triangles are metal-poor, metal-rich and anomalous stars,
respectively, from \citet{Johnson:2015aa}. 
      \label{fig:cmd} }
\end{figure}

Program stars (see Table \ref{tab:stars}) were selected from $JK$
colour-magnitude diagrams (CMD) using 2MASS photometry
\citep{Skrutskie:2006aa}. All targets occupy plausible locations in the $J-K$
vs.\ $K$ CMD. Specifically, we required 0.68 $\le$ $J-K$ $\le$ 1.2 and
8.57\footnote{This value corresponds to the tip of the RGB
\citep{Valenti:2007aa}.} $\le$ $K$ $\le$ 13.0 (see Figure \ref{fig:cmd}).
Furthermore, all program stars were required to have ``AAA'' 2MASS photometric
quality flags and to lie within 3 arcmin of the cluster centre. 

\begin{table*}
 \centering
 \begin{minipage}{190mm}
  \caption{Program stars.} 
  \label{tab:stars} 
  \begin{tabular}{@{}lcccrrr@{}}
  \hline
        Name (2MASS) & 
	RA (J2000) &
	Dec (J2000) &
	Mask &
	$J$ & 
	$H$ & 
	$K$ \\ 
\hline 
  \hline
17022878-2614320 &  255.61995 & $-$26.24223 &  1 & 11.304 & 10.556 & 10.393 \\ 
17023192-2614177 &  255.63304 & $-$26.23827 &  1 & 10.451 &  9.666 &  9.434 \\ 
17023394-2616196 &  255.64145 & $-$26.27213 &  1 & 10.444 &  9.725 &  9.470 \\ 
17023460-2616038 &  255.64418 & $-$26.26775 &  1 & 11.581 & 10.897 & 10.699 \\ 
17023481-2617152 &  255.64505 & $-$26.28756 &  1 & 11.415 & 10.681 & 10.535 \\ 
 \hline
\end{tabular}
\end{minipage}
This table is published in its entirety in the electronic edition of the paper.
A portion is shown here for guidance regarding its form and content. 
\end{table*}

Observations of the program stars were obtained using the GMOS-S multi-object
spectrograph \citep{Hook:2004aa} at the Gemini-S telescope in queue mode
(GS-2012A-Q-58). All observations were taken on 2012 May 05 and the set-up was
identical to that of \citet{Da-Costa:2014aa}. We used the R831 grating and the
RG610 filter with central wavelengths of 8550\,\AA\ and 8600\,\AA\ (wavelength
coverage was typically from $\sim$7500\,\AA\ to $\sim$9000\,\AA). Since there
are gaps between the three GMOS-S CCDs, observations at two different central
wavelengths ensures that the \caii\ infrared triplet lines can be measured in
at least one of the two settings. Each GMOS mask had slit widths of 1 arcsec
and slit lengths of 10 arcsec to facilitate sky and background subtraction. A
total of 92 candidate RGB stars were observed across six masks. Two stars
(17024132-2613517 and 17024717-2615107) were common to two masks and a total of
21 stars from J15 were included across five masks. Masks 1, 2 and 3
concentrated on brighter targets ($J$ $\le$ 11.7) and the total exposure time
per mask was 240 sec (one 120 sec exposure for each of the two central
wavelengths). Masks 4, 5 and 6 concentrated on fainter targets ($J$ $\ge$ 11.5)
and the total exposure time per mask was 960 sec (one 480 sec exposure for each
central wavelength). Each observation was preceded, or followed, by a
flat-field integration. Arc lamp exposures for each mask were also obtained. 

\begin{figure}
\centering
      \includegraphics[width=.99\hsize]{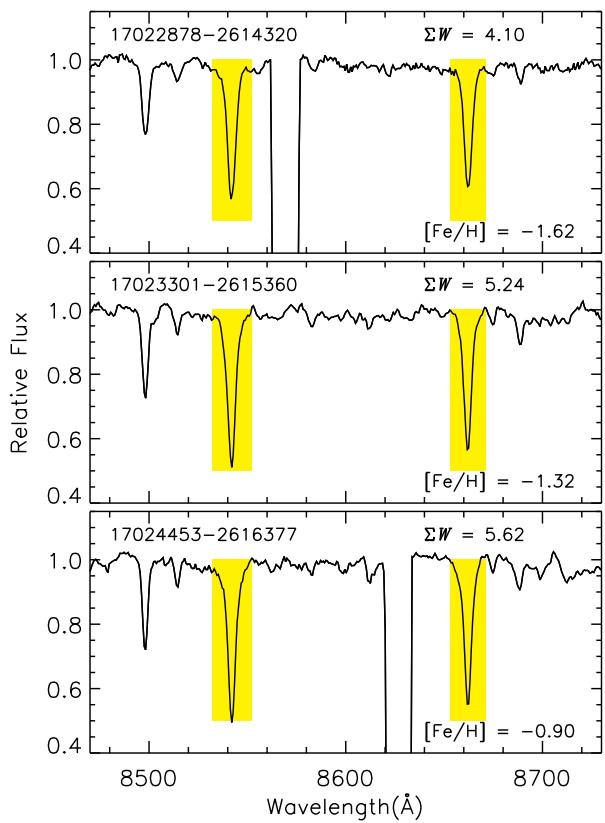} 
      \caption{A portion of the spectra with central wavelength 8550\,\AA\ for
three stars with similar $K$ magnitudes, all studied by J15, ordered by
increasing metallicity. The positions of the 8542\,\AA\ and 8662\,\AA\ \caii\
lines are indicated in yellow. The sum of the equivalent widths, $\Sigma W$, is
shown in each panel along with the final metallicity. In the upper and lower
panels, the regions with zero flux are due to gaps between the CCDs. 
      \label{fig:spec} }
\end{figure}

\begin{figure}
\centering
      \includegraphics[width=.99\hsize]{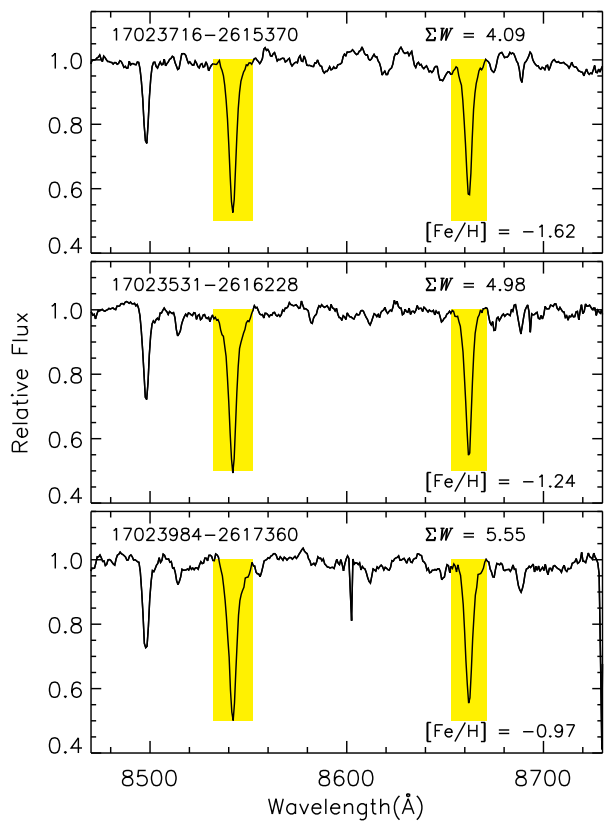} 
      \caption{Same as Figure \ref{fig:spec} but for three different (fainter)
stars with central wavelength 8600\,\AA. 
      \label{fig:spec2} }
\end{figure}

Data reduction was performed using {\sc iraf}\footnote{{\sc iraf} is
distributed by the National Optical Astronomy Observatories, which are operated
by the Association of Universities for Research in Astronomy, Inc., under
cooperative agreement with the National Science Foundation.} Gemini package
scripts. The final wavelength-calibrated, sky-subtracted spectra have a
resolution of $\sim$3.5\,\AA\ and a (binned) pixel scale of 0.68\,\AA\ per
pixel (see Figures \ref{fig:spec} and \ref{fig:spec2}). The signal-to-noise
ratios (S/N), based on the photon counts for a given central wavelength, exceed
100 per pixel for all but three objects with an average value of 170 and a
maximum value of 280. 

As noted by \citet{Da-Costa:2014aa}, the GMOS-S CCDs (at the time of the
observations) are affected by strong fringing longward of $\sim$7000\,\AA. The
main purpose of these spectra is to measure radial velocities and metallicities
based on the strength of the \caii\ triplet features. Therefore, even though
the spectra have high S/N, the accuracy of the equivalent width measurements is
limited by the large amplitude fringing rather than the photon counts (see
Figures \ref{fig:spec} and \ref{fig:spec2}). Consequently, we treated the
spectra taken at the two different central wavelengths independently rather
than co-adding them. For spectra of the same star taken with different central
wavelengths, the calcium triplet lines will fall at different locations on the
CCD and be subject to different fringing patterns. As we shall describe later,
for each star we will compare the equivalent width measurements from the two
spectra and select only those stars for which there is good agreement. 

Radial velocities were measured from the wavelengths of the three calcium
triplet lines. In a given spectrum, the radial velocity from the three lines
were in good agreement and were averaged. For a given star, the radial
velocities from the spectra taken with different central wavelengths
(8550\,\AA\ and 8600\,\AA) were also in good agreement (maximum absolute
difference of 3.7 \kms\ between the two spectra) and were averaged to produce
the final radial velocity. 

The zero-points for the radial velocity measurements, however, are not reliable
because the arc lamp exposures were taken during the daytime (and for Mask 1,
the arc lamp exposures were taken two days after the science observations). Two
stars, 17024132-2613517 and 17024717-2615107, were observed in Masks 1 and 4
and the radial velocities differed by 5.5 and 15.3 \kms, respectively, between
the two masks. Fortunately, for Masks 1 through 5, we had six, five,
five, four and one star, respectively, in common with J15. By comparing our
radial velocities with those of J15, we could place our measurements onto their
scale. The minimum and maximum velocity shifts applied to a given mask were
$-$7.2 and +5.7 \kms, respectively. For Mask 6, however, there were no stars
in common with J15 and so we shifted the velocities in Mask 6 by +16.4 \kms\ to
match the average value obtained from Masks 1 through 5. Therefore, we regard
the uncertainty in any given radial velocity measurement to be $\sim$15 \kms. 

\begin{figure}
\centering
      \includegraphics[width=.99\hsize]{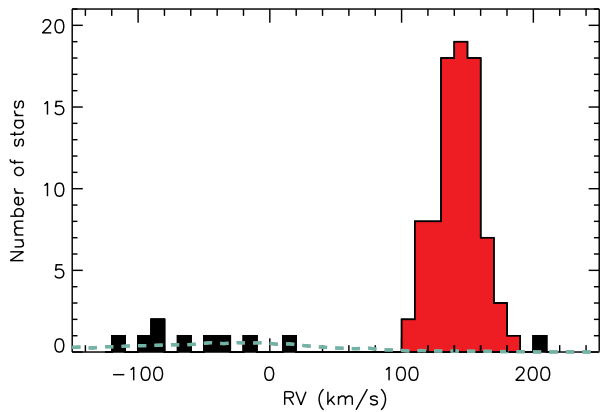} 
      \caption{Distribution of radial velocities. Stars outside the radial
velocity range 100-190 \kms\ were regarded as non-members. The aqua dashed line
is the predicted distribution of field stars from the Besan\c{c}on model
\citep{Robin:2003aa}, normalised to the number of stars with RV $<$ 100 \kms. 
      \label{fig:rv} }
\end{figure}

In Figure \ref{fig:rv}, we plot the distribution of radial velocities for the
program stars. Most of the data lie near +145 \kms\ and we reject stars with RV
$\le$ +100 \kms\ and RV $\ge$ +190 \kms. For the stars we regard to be radial
velocity members, we find an average radial velocity of +144.1 \kms\ (the
average value in J15 is +144.5 \kms). While the radial velocity measurements
(see Table \ref{tab:param}) can help to identify non-members, we cannot obtain
a meaningful velocity dispersion from our GMOS-S spectra. 

\begin{table*}
 \centering
 \begin{minipage}{190mm}
  \caption{Radial velocities, equivalent widths and metallicities.} 
  \label{tab:param} 
  \begin{tabular}{@{}lcccccccccrc@{}}
  \hline
        Name (2MASS) & 
	Mask &
	RV & 
	EW$_{\rm 8542}$ & 
	EW$_{\rm 8542}$ & 
	EW$_{\rm 8662}$ & 
	EW$_{\rm 8662}$ & 
	$\Sigma W$ & 
	$\sigma\Sigma W$~ & 
	$W^\prime$ & 
	[Fe/H] &
	$\sigma$[Fe/H]~ \\
        & 
	&
	& 
	(8550\,\AA)& 
	(8600\,\AA)& 
	(8550\,\AA)& 
	(8600\,\AA)& 
	& 
	& 
	& 
	&
	\\ 
        & 
	&
	\kms & 
	\AA& 
	\AA& 
	\AA& 
	\AA& 
	\AA & 
	\AA & 
	\AA & 
	dex &
	dex 
 \\
\hline 
  \hline
17022878-2614320 &  1 &   144 &    2.24 &    2.65 &    1.86 &    1.88 &    4.32 &    0.21 &    3.57 & $-$1.62 &    0.09 \\ 
17023192-2614177 &  1 &   129 &    2.83 &    3.16 &    2.24 &    1.85 &    5.05 &    0.26 &    3.97 & $-$1.45 &    0.13 \\ 
17023394-2616196 &  1 &   155 &    2.70 &    2.50 &    1.75 &    1.92 &    4.44 &    0.13 &    3.34 & $-$1.71 &    0.06 \\ 
17023460-2616038 &  1 &   164 &    2.54 &    2.47 &    1.64 &    2.08 &    4.37 &    0.22 &    3.74 & $-$1.55 &    0.10 \\ 
17023481-2617152 &  1 &   155 &    3.26 &    2.72 &    2.03 &    1.79 &    4.90 &    0.30 &  \ldots &  \ldots &  \ldots \\ 
 \hline
\end{tabular}
\end{minipage}
This table is published in its entirety in the electronic edition of the paper.
A portion is shown here for guidance regarding its form and content. 
\end{table*}

To estimate field contamination, we used the Besan\c{c}on model
\citep{Robin:2003aa}.  We considered all stars within one square degree
centered on NGC 6273 and only accepted stars with 0.68 $\le$ $J-K$ $\le$ 1.2
and 8.57 $\le$ $K$ $\le$ 13.0.  In Figure \ref{fig:rv}, we overplot the
predicted distribution of field stars, normalised to the number of program
stars with RV $<$ 100 \kms, i.e., non-members. We therefore expect minor
contamination from field stars, only $\sim$1 star with +100 $\le$ RV $\le$ +200
\kms. 

\begin{figure}
\centering
      \includegraphics[width=.99\hsize]{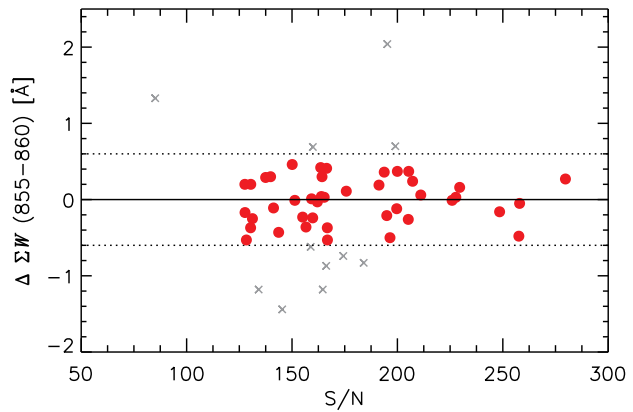} 
      \caption{Difference in $\Sigma W$ between the 8550\,\AA\ and 8600\,\AA\
spectra versus S/N for each program star for which measurements of EW$_{\rm
8542}$ and EW$_{\rm 8662}$ could be obtained from both sets of spectra. Filled
red circles are stars with $|\Delta \Sigma W|$ $\le$ 0.6\,\AA. 
      \label{fig:caterrsn} }
\end{figure}

Equivalent widths of the calcium triplet lines were measured by fitting Voigt
profiles using routines in {\sc iraf} adopting the definitions in
\citet{Armandroff:1991aa}. For a given star, we could measure the strengths of
the 8542\,\AA\ and 8662\,\AA\ \caii\ lines in the spectra taken with central
wavelengths 8550\,\AA\ and 8600\,\AA\ (see Table \ref{tab:param}). For a given
central wavelength, we could then add the equivalent widths for the 8542\,\AA\
and 8662\,\AA\ lies to produce $\Sigma W$. For each star, we can plot the
difference in $\Sigma W$ between the spectra taken with central wavelengths
8550\,\AA\ and 8600\,\AA\ as a function of S/N (see Figure \ref{fig:caterrsn}).
Only stars for which we were able to measure both lines in both spectra are
included in this figure (and we have excluded radial velocity non-members). As
in \citet{Da-Costa:2014aa}, while there is reasonable agreement in the two
measurements of $\Sigma W$ for most stars, the difference can be substantial
for some stars and persists despite the very high S/N. We attribute these
differences to fringing. We arbitrarily adopt a cutoff of $|\Delta$\,$\Sigma
W|$ = 0.6\,\AA\ and S/N $\ge$ 100 in order to select only the best spectra. For
these 42 stars, the average difference in $\Sigma W$ between the 8550\,\AA\ and
8600\,\AA\ spectra is $-$0.01 $\pm$ 0.05\,\AA\ ($\sigma$ = 0.29). For these
stars, we average the measurements from the 8550\,\AA\ and 8600\,\AA\ spectra
to obtain the final $\Sigma W$ (see Table \ref{tab:param}). 

\section{RESULTS} 

\begin{figure*}
\centering
      \includegraphics[width=.80\hsize]{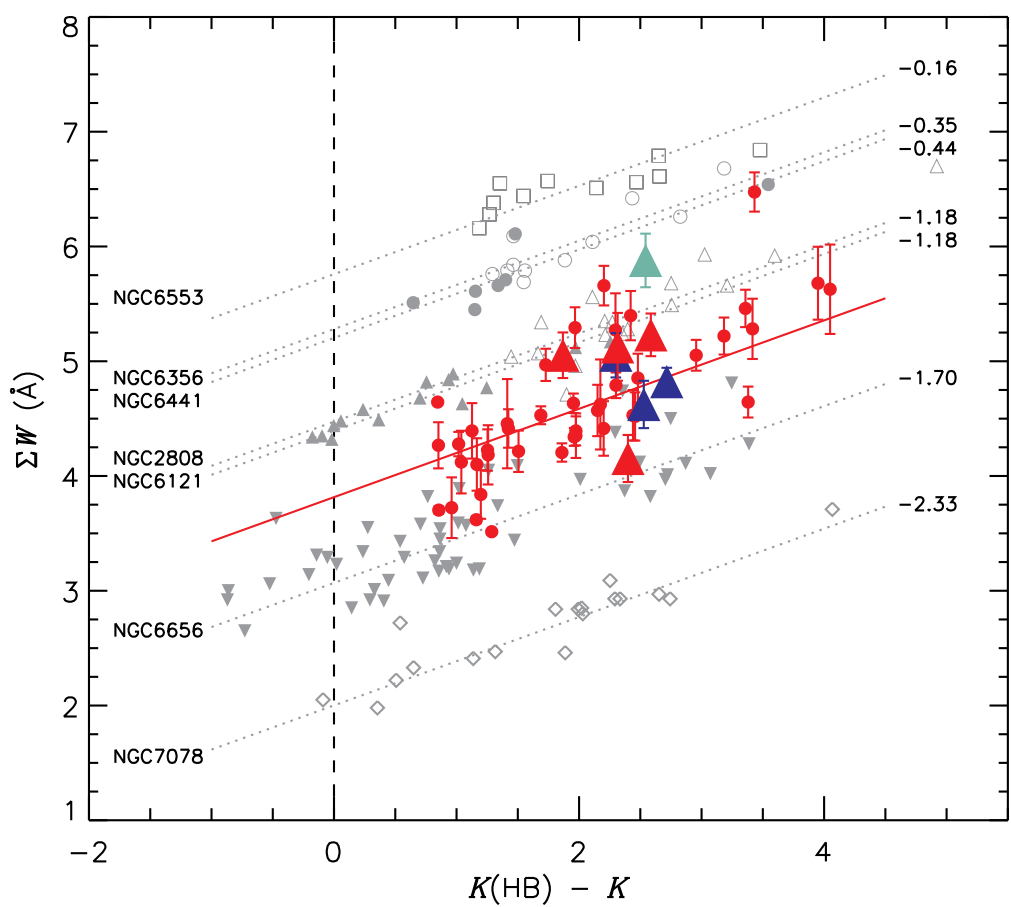} 
      \caption{Sum of the 8542\,\AA\ and 8662\,\AA\ equivalent widths
($\Sigma W$) as a function of the magnitude difference from the horizontal
branch for 42 program stars (red filled circles). The red solid line is the
average value when imposing a gradient of $-$0.385\,\AA/mag. The large
triangles are eight stars with metallicity measurements from J15. Comparison
globular clusters (individual stars and the best fitting line) from
\citet{Mauro:2014aa} are overplotted. 
      \label{fig:cat} }
\end{figure*}

In Figure \ref{fig:cat}, we plot the sum of the 8542\,\AA\ and 8662\,\AA\
equivalent widths ($\Sigma W$) as a function of the magnitude difference from
the horizontal branch for 42 stars as filled red circles. We applied a
line-strength correction factor of 1.046 following \citet{Da-Costa:2014aa} and
adopted $K$(HB) = 12.85 from \citet{Valenti:2007aa}. We also include
comparison globular clusters from \citet{Mauro:2014aa}. Of the 21 stars in
common with J15, eight have metallicities from J15 and we overplot those
objects as large filled triangles. Of those eight objects, three are from the
``metal-poor'' population (blue), four are from the ``metal-rich'' population
(red) and the sole ``anomalous'' star is included (aqua). We caution that for
five of these eight stars in common with J15, we were unable to measure
equivalent widths for the 8542\,\AA\ and 8662\,\AA\ lines in both spectra
(8550\,\AA\ and 8600\,\AA). Therefore we were unable to compare $\Sigma W$
between the 8550\,\AA\ and 8600\,\AA\ spectra and so the $\Sigma W$
measurements for those five stars may have larger errors than for the 42
program stars. We will comment on these stars later in this section. 

Following \citet{Mauro:2014aa}, we use the $K$ magnitude rather than the $V$
magnitude in Figure \ref{fig:cat}. The advantages to this approach are that
extinction and differential reddening in the $K$ band are much lower when
compared to the $V$ band. Additionally, the relatively flat slope of the
$\Sigma W$ vs.\ $K$(HB)$-K$ relation ($-$0.385\,\AA/mag compared to
$-$0.63\,\AA/mag for the optical relation) means that errors arising from
photometry, distance and/or reddening will have a smaller impact upon the
derived metallicities. 

To obtain metallicities, we adopted the approach of \citet{Mauro:2014aa} and
assumed the relation $\Sigma W$ = 0.385[$K$(HB) $-$ $K$] + $W^\prime$. For each
star, we can then obtain the reduced equivalent width, $W^\prime$. For the 42
stars shown in Figure \ref{fig:cat}, we find an average reduced equivalent
width of $\langle W^{\prime} \rangle$ = 3.86 $\pm$ 0.07\,\AA\ ($\sigma$ =
0.42\,\AA). Note that the standard deviation of 0.42\,\AA\ cannot be attributed
entirely to the difference in $\Sigma W$ measured from the two different
settings, 8550\,\AA\ and 8600\,\AA, nor to the uncertainties in $\Sigma W$ (the
average uncertainty is 0.19 $\pm$ 0.01\,\AA). Therefore, there is an intrinsic
spread in $\Sigma W$, i.e., \caii\ line strengths, in NGC 6273. 

\begin{figure}
\centering
      \includegraphics[width=.99\hsize]{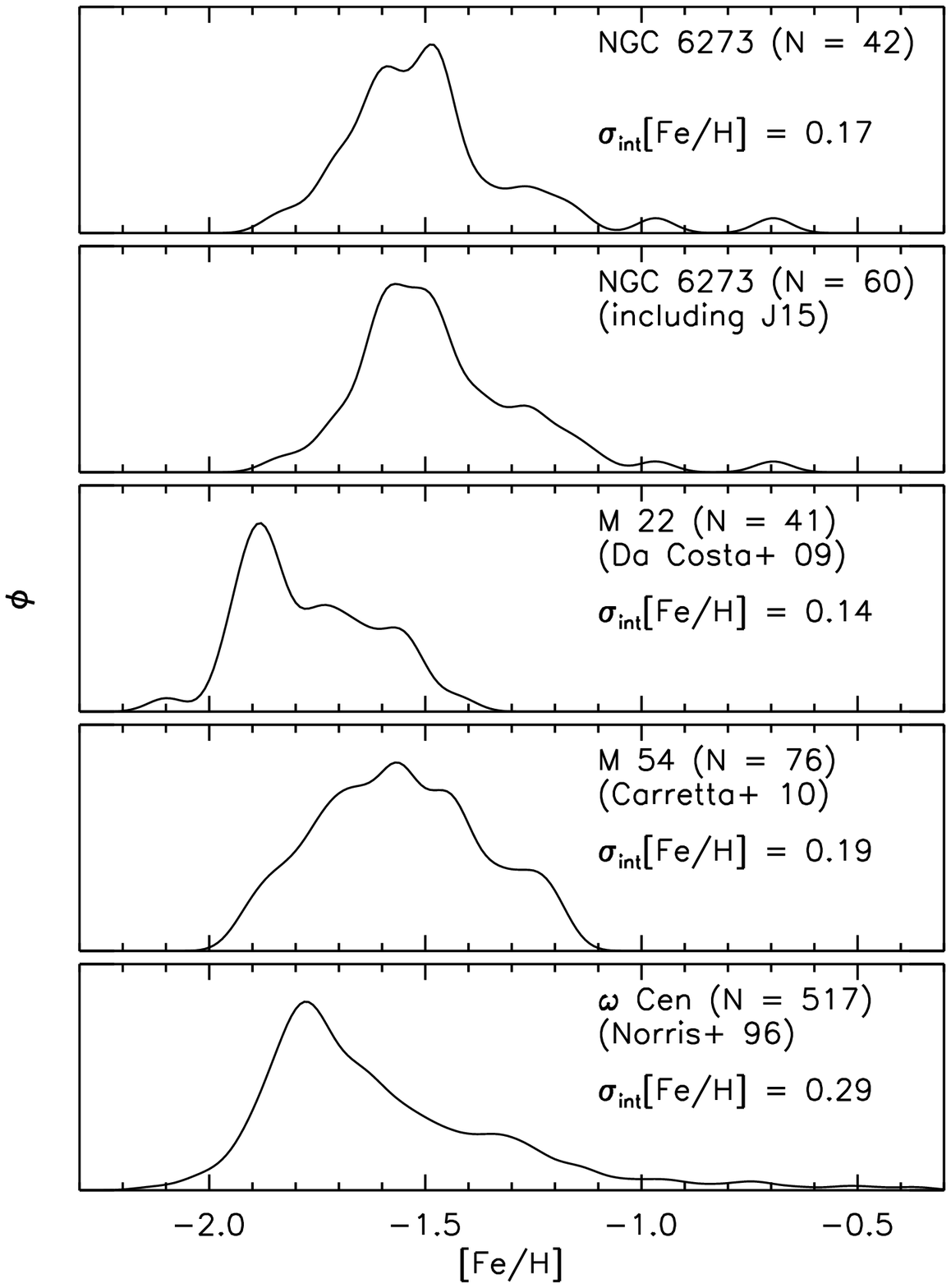} 
      \caption{Abundance distributions (in generalised histogram form) for NGC
6273 (upper two panels), M 22 (third panel, data from
\citealt{Da-Costa:2009aa}), M 54 (fourth panel, data from
\citealt{Carretta:2010ab}) and $\omega$ Cen (lower panel, data from
\citealt{Norris:1996aa}, see text for details). The second panel includes all
stars from J15 but with metallicities shifted by +0.20 dex (see text for
details). The intrinsic dispersion is written in each panel (see text for
details) as well as the number of stars. 
      \label{fig:feh} }
\end{figure}

To convert the reduced equivalent widths into metallicities, we use the
following relation from \citet{Mauro:2014aa} which is calibrated by the
observed line strengths in standard globular clusters with metallicities on the
\citet{Carretta:2009ab} abundance scale: [Fe/H] = $-$4.61 + 1.842$\langle
W^{\prime} \rangle$ $-$ 0.4428$\langle W^{\prime} \rangle$$^2$ +
0.04517$\langle W^{\prime} \rangle$$^3$. Using this calibration, we obtain an
average metallicity of [Fe/H] = $-$1.48 $\pm$ 0.03 ($\sigma$ = 0.21). We plot
the metallicity distribution in Figure \ref{fig:feh}, upper panel, as a
generalised histogram in which each data point is represented by a unit
Gaussian of width 0.10 dex. (The average error in [Fe/H] is close to 0.10 dex). 

We now seek to quantify the abundance dispersion. For the 42 program stars, the
standard deviation for $W^\prime$ is 0.42\,\AA\ and the average error is
0.19\,\AA. When taking into account the error, the intrinsic spread in
$W^\prime$ is therefore $\sim$0.38\,\AA. At the metallicity of NGC 6273, that
intrinsic spread in reduced equivalent width translates into an intrinsic
abundance spread of $\sigma_{\rm int}$[Fe/H] = 0.17 dex. We therefore confirm
the results of J15 who reported a very similar value of $\sigma$[Fe/H] = 0.16
dex and an average metallicity of [Fe/H] = $-$1.62. 

In order to select stars with reliable calcium triplet line strength
measurements, we required $\Sigma W$ to be in good agreement between the
8550\,\AA\ and 8600\,\AA\ observations; $|\Delta$\,$\Sigma W|$ $\le$ 0.6\,\AA\
as seen in Figure \ref{fig:caterrsn}.  Had we adopted a more stringent cutoff
of $|\Delta$\,$\Sigma W|$ $\le$ 0.4\,\AA, our results and conclusions would be
essentially unchanged: for the 34 stars that meet that criterion, the standard
deviation in $W^\prime$ is 0.46\,\AA\ and the average error in $\Sigma W$ is
0.17\,\AA, i.e., the dispersion in $W^\prime$ cannot be explained by errors in
$\Sigma W$ alone. Instead, the intrinsic spread in $W^\prime$ is 0.42\,\AA\ and
this translates into an intrinsic abundance spread of $\sigma_{\rm int}$[Fe/H]
= 0.19 dex. 

NGC 6273 suffers from large and variable reddening, $E(B-V)$ = 0.32 and $\Delta
E(B-V)$ $\sim$ 0.3 \citep{Alonso-Garcia:2012aa}. One possibility is that
differential reddening is responsible (in part or in whole) for the dispersion
in $W^\prime$ and metallicity dispersion. To check this possibility, we used 
the differential reddening map for NGC 6273 from \citet{Alonso-Garcia:2012aa}.
For each program star, we averaged the differential reddening values within 5
arcsec. The average value was $\Delta E(B-V)$ = 0.015 $\pm$ 0.006 ($\sigma$ = 0.068)
with minimum and maximum values of $-$0.100 and 0.174. Assuming $A_K$ =
0.34$E(B-V)$, the standard deviation in $W^\prime$ is unchanged and therefore,
differential reddening as high as $\Delta E(B-V)$ = $\sim$0.27 does not affect the
derived metallicities. As noted, using the $K$ magnitude rather than the $V$
magnitude greatly reduces the effect of differential reddening and the slope of
the $\Sigma W$ vs.\ $K$(HB)$-$$K$ relation is flatter than the corresponding
optical trend. 

Recall that we also observed eight stars with metallicities from J15. For five
of these objects, however, we could not measure the equivalent widths for
8542\,\AA\ and 8662\,\AA\ in both spectra. While their $\Sigma W$ measurements
may have larger uncertainties, we can obtain metallicities and compare our
values with J15 (see Table \ref{tab:j15}). The average difference is
$\Delta$[Fe/H] = 0.20 $\pm$ 0.07 ($\sigma$ = 0.20 dex), with our values being
higher than those of J15. On our abundance scale, the three ``metal-poor''
stars and four ``metal-rich'' stars from J15 have average metallicities of
[Fe/H] = $-$1.49 and [Fe/H] = $-$1.42, respectively. The ``anomalous'' star has
[Fe/H] = $-$0.90. In the second panel of Figure \ref{fig:feh}, we combine our
program stars with all stars from J15 but we shift their metallicities by +0.20
dex to match our scale. The upper two panels exhibit very similar abundance
distributions. That is, including or excluding the J15 stars does not affect
our results, namely, that there is an intrinsic abundance dispersion in NGC
6273. 

\begin{table}
 \centering
 \begin{minipage}{85mm}
  \caption{Equivalent widths and metallicities for eight stars from J15.} 
  \label{tab:j15} 
  \begin{tabular}{@{}lccccc@{}}
  \hline
        Name (2MASS) & 
	$W^\prime$ & 
	$\sigma W^\prime$~ & 
	[Fe/H] &
	$\sigma$[Fe/H] &
	[Fe/H] \\
         & 
	 & 
	 & 
	 & 
	 &
	(J15) \\ 
         &
	\AA & 
	\AA & 
	dex & 
	dex &
	dex 
 \\
\hline 
  \hline
\multicolumn{6}{c}{Metal-poor population} \\
17024618-2615261 & 3.65 & 0.21 & $-$1.59 & 0.09 & $-$1.76 \\
17023856-2617209 & 3.78 & 0.12 & $-$1.54 & 0.05 & $-$1.80 \\
17023158-2617259 & 4.17 & 0.19 & $-$1.35 & 0.10 & $-$1.80 \\
\multicolumn{6}{c}{Metal-rich population} \\
17023481-2617152 & 4.23 & 0.30 & $-$1.32 & 0.16 & $-$1.44 \\
17023301-2615360 & 4.23 & 0.19 & $-$1.32 & 0.11 & $-$1.37 \\
17024016-2616096 & 3.23 & 0.21 & $-$1.76 & 0.08 & $-$1.60 \\
17024326-2617504 & 4.33 & 0.20 & $-$1.27 & 0.12 & $-$1.55 \\
\multicolumn{6}{c}{Anomalous population} \\
17024453-2616377 & 4.90 & 0.23 & $-$0.90 & 0.19 & $-$1.30 \\
 \hline
\end{tabular}
\end{minipage}
\end{table}

Of particular interest is the possibility that NGC 6273 hosts a third,
``anomalous'', population. J15 identified one ``anomalous'' star,
17024453-2616377, with [Fe/H] = $-$1.30 on their abundance scale. That star has
[Fe/H] = $-$0.90 on our abundance scale. We identify four additional stars with
[Fe/H] $>$ $-$1.2: 17023649-2615229 ([Fe/H] = $-$1.19), 17023685-2616217
([Fe/H] = $-$1.15), 17023984-2617360 ([Fe/H] = $-$0.97) and 17023595-2615342
([Fe/H] = $-$0.70). Assuming that at least one of those stars is a genuine
cluster member (recall that the Besan\c{c}on model  predicted only $\sim$1
field star contaminant), then our spectra confirm that NGC 6273 hosts an
``anomalous'' population, i.e., a high metallicity tail. 

\begin{figure}
\centering
      \includegraphics[width=.99\hsize]{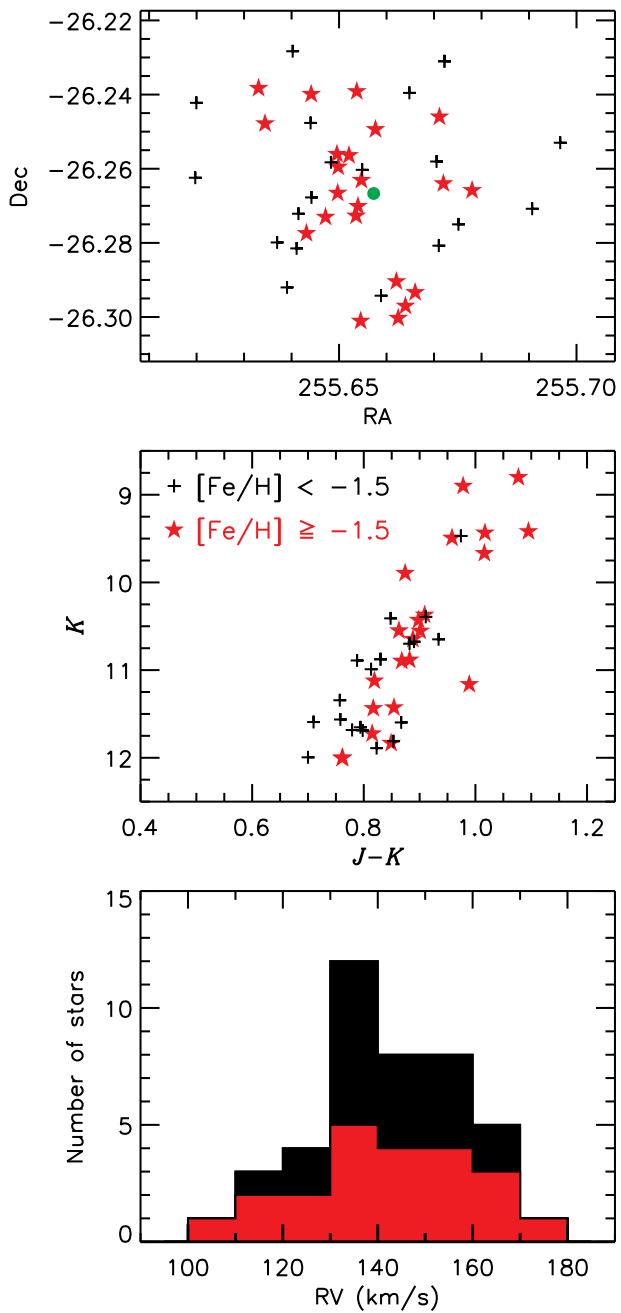} 
      \caption{Spatial distribution (upper), CMD (middle) and radial velocity
distribution (lower). Black crosses and red stars represent objects with [Fe/H]
$<$ $-$1.5 and [Fe/H] $\ge$ $-$1.5, respectively. 
      \label{fig:mr} }
\end{figure}

In Figure \ref{fig:mr}, we separate the program stars based on their
metallicity into two groups to investigate whether there are any systematic
differences in their spatial distributions, location in the CMD or radial
velocity distributions. We arbitrarily choose [Fe/H] = $-$1.50 as the boundary
between the two groups. From this figure, the spatial and radial velocity
distributions for the two populations overlap. Regarding the CMD, while we
would expect the more metal-rich stars to be redder at a given $K$ magnitude
when compared to the more metal-poor stars, we find no significant differences
in colour between the two populations. 

\section{DISCUSSION} 

We confirm the results from J15 that NGC 6273 harbours an internal abundance
dispersion. In the lower three panels of Figure \ref{fig:feh}, we also plot the
metallicity distributions for well-studied globular clusters that also exhibit
abundance dispersions, M 22 \citep{Da-Costa:2009aa}, M 54
\citep{Carretta:2010ab} and $\omega$ Cen \citep{Norris:1996aa}\footnote{For the
$\omega$ Cen data, \citet{Norris:1996aa} calibrate the \caii\ line strengths to
[Ca/H], rather than [Fe/H]. \citet{Norris:1995aa} find a mean [Ca/Fe] of +0.39,
and we have shifted the \citet{Norris:1996aa} values by $-$0.39 dex to obtain 
[Fe/H].}. The same smoothing, FWHM = 0.10 dex, has been applied to all panels
in this figure. For M 22 and $\omega$ Cen, the metallicity measurements were
made using intermediate resolution spectra of the \caii\ triplet lines, i.e., a
similar approach as in this study. For M 54, however, the metallicities are
obtained from analysis of individual Fe lines based on high-resolution spectra. 

In the upper panel of Figure \ref{fig:feh}, we write the intrinsic abundance
dispersion for NGC 6273 which was computed and described in the previous
Section. For M 22 \citep{Da-Costa:2009aa}, M 54 \citep{Carretta:2010ab} and
$\omega$ Cen \citep{Norris:1996aa}, the measurement errors are 0.06, 0.02, and
0.05 dex, respectively. To obtain the intrinsic abundance dispersion,
$\sigma_{\rm int}$[Fe/H], we assume that the observed standard deviation is the
quadratic sum of the measurement error and the intrinsic abundance dispersion,
and we write the latter quantity in the bottom three panels. NGC 6273 has
$\sigma_{\rm int}$[Fe/H] = 0.17 dex which is marginally smaller than that of M
54, $\sigma_{\rm int}$[Fe/H] = 0.19 dex. M 22 has a slightly smaller value,
$\sigma_{\rm int}$[Fe/H] = 0.14, while $\omega$ Cen has the highest value,
$\sigma_{\rm int}$[Fe/H] = 0.29 dex. 

At low metallicity, the abundance distribution of NGC 6273 rises rapidly in a
similar manner as $\omega$ Cen, but perhaps not as steeply as M 22. For M 54,
the abundance distribution appears to rise more slowly compared to the other
globular clusters. At high metallicity, NGC 6273 has a tail that extends to at
least [Fe/H] = $-$1.0, and perhaps as high as [Fe/H] = $-$0.70. Overall, the
shape of the abundance distribution of NGC 6273 most closely resembles that of
$\omega$ Cen. 

We note that \citet{Mucciarelli:2015aa} have challenged the existence of an
iron abundance dispersion in M 22 reported by
\citet{Marino:2009aa,Marino:2011ab} based on high-resolution spectroscopy.
\citet{Mucciarelli:2015aa} found that when using surface gravities based on
photometry, the [\feii/H] distribution exhibits no spread. They suggested that
neglect of non-local thermodynamic equilibrium (NLTE) effects can produce
spurious abundance spreads. For M 22, however, there is also a real spread in
\caii\ line strengths \citep{Da-Costa:2009aa} which, if not due to metallicity
variations, requires an unknown explanation. Recent NLTE calculations for Fe
\citep{Bergemann:2012aa,Lind:2012aa} indicate that for stars with similar
stellar parameters, the NLTE corrections are very similar (i.e., the
differential NLTE effects are very small). We consider the abundance spreads in
M 22 as real given the concordance between the high dispersion results and the
\caii\ results for this cluster and that the differential NLTE corrections are
expected to be negligible. 

The increase in $s$-process element abundances with increasing metallicity is a
characteristic shared by $\omega$ Cen, M 22, M 2, NGC 5286 and NGC 6273 and
requires contributions from Type II supernovae and AGB stars. The presence of a
third, ``anomalous'', population at higher metallicity but without $s$-process
element enhancements or light element abundance anomalies has only been
identified in M 2 (four stars) and NGC 6273 (one star). Clearly, it would be of
great interest to examine chemical abundances for a larger suite of elements in
the most metal-rich stars of NGC 6273 to establish whether the anomalous
populations in these two globular clusters share common characteristics. 

It is worth noting that initial observations of the \caii\ triplet in the outer
halo globular cluster NGC 2419 showed line strength variations
\citep{Cohen:2010aa}. Subsequent analyses, however, revealed that this cluster
has a single metallicity, [Fe/H] = $\sim-$2.1, but is highly unusual in that it
exhibits very large star-to-star variations for Mg and K (factors of $\sim$100)
\citep{Cohen:2011aa,Cohen:2012aa,Mucciarelli:2012aa}. Mg is an important
electron donor in low mass stars (and thus contributes to the H$^-$ continuous
opacity), and the large Mg depletion combined with a small star-to-star Ca
variation could explain the behaviour of the \caii\ triplet lines in NGC 2419
\citep{Cohen:2012aa,Mucciarelli:2012aa}. A similar explanation could possibly
also apply to NGC 5824 \citep{Da-Costa:2014aa,Roederer:2016aa}. For NGC 6273,
however, the calcium line strength variation is larger than in NGC 2419 and NGC
5824, the Mg variation is relatively modest ($\Delta$[Mg/Fe] $\simeq$ 0.5; J15)
and the overall metallicity is higher. So it seems unlikely that star-to-star
Mg variations are driving the calcium line strength variations in NGC 6273. 

We reiterate that the most plausible explanation for the origin of $\omega$ Cen
is that it is the surviving nucleus of an accreted dwarf galaxy
\citep{Freeman:1993aa,Lee:1999aa,Bekki:2003aa}. It has been argued that when
the Sgr dSph is tidally disrupted by the Milky Way, the remnant of the M 54+Sgr
system may closely resemble $\omega$ Cen \citep{Carretta:2010ac}. Since
$\omega$ Cen, M 54 and NGC 6273 all exhibit an intrinsic abundance dispersion,
it is tempting to suggest that they may all share a similar formation mechanism
as the remnant nuclei of disrupted dwarf galaxies. If this is the case, then
one might expect to find a diffuse stellar envelope (and/or extratidal stars)
surrounding NGC 6273 as is the case for other globular clusters that exhibit
abundance dispersions in Fe-peak and/or $s$-process elements including NGC 1851
\citep{Olszewski:2009aa,Marino:2014aa,Navin:2015aa}, NGC 5824
\citep{Grillmair:1995aa}, M 2 (\citealt{Grillmair:1995aa}, Kuzma et al.\ in
prep) and M 22 \citep{Kunder:2014aa}. Identifying such a stellar halo will be
challenging given that NGC 6273 lies near the Galactic bulge and has large and
variable reddening, $E(B-V)$ = 0.32 and $\Delta E(B-V)$ $\sim$ 0.3
\citep{Alonso-Garcia:2012aa}. 

Finally, two ($\omega$ Cen and M 22) of the clusters previously known to
exhibit variations of the heavy elements Ca and/or Fe are also known to exhibit
very large variations in cyanogen and to exhibit large ellipticities (see
\citealt{Norris:1987aa})\footnote{The fact that the third cluster, M 54, has
only a small measured ellipticity ($\epsilon \sim 0.06$,
\citealt{White:1987aa}) may result from projection effects.}. We note then that
NGC 6273 exhibits the largest ellipticity of the Galactic globular cluster
population ($\epsilon \sim 0.28$, \citealt{White:1987aa}). From narrowband
Ca photometry, \citet{Han:2015aa} identified a Ca-rich and Ca-poor population
in NGC 6273 and measured the CN and CH line strength indices in each
population. They found that the Ca-rich population was also enhanced in CN,
which is similar to $\omega$ Cen and M 22 \citep{Norris:1983aa}. 

\section{CONCLUSIONS} 

We confirm that the globular cluster NGC 6273 harbours an intrinsic abundance
spread based on intermediate resolution GMOS-S spectra of the line strengths of
the \caii\ triplet features. The intrinsic abundance dispersion, $\sigma_{\rm
int}$[Fe/H] = 0.17 agrees with the results from J15, $\sigma$[Fe/H] = 0.16. NGC
6273 therefore has the third largest abundance dispersion after $\omega$ Cen
and M 54 among the Galactic globular clusters. The total range in metallicity
is $\sim$1.0 dex. We confirm the presence of a high metallicity tail with
values reaching above [Fe/H] = $-$1.0 and perhaps up to [Fe/H] = $-$0.70. Given
the similarities in the abundance distribution, we argue that NGC 6273 and
$\omega$ Cen may share the same origin as the nuclei of disrupted dwarf
galaxies. 

\section*{Acknowledgments}

We thank the anonymous referee for a careful reading of the paper and helpful
comments. 
We gratefully acknowledge support from the Australian Research Council (grants
DP0984924, DP120101237, DP120100475, FT140100554 and DP150103294).

\label{lastpage}


\begin{thebibliography}{64}
\expandafter\ifx\csname natexlab\endcsname\relax\def\natexlab#1{#1}\fi

\bibitem[{{Alonso-Garc{\'{\i}}a} {et~al}\mbox{.}(2012){Alonso-Garc{\'{\i}}a},
  {Mateo}, {Sen}, {Banerjee}, {Catelan}, {Minniti}, \& {von
  Braun}}]{Alonso-Garcia:2012aa}
{Alonso-Garc{\'{\i}}a} J., {Mateo} M., {Sen} B., {Banerjee} M., {Catelan} M.,
  {Minniti} D., {von Braun} K., 2012, \aj, 143, 70

\bibitem[{{Armandroff} \& {Da Costa}(1991)}]{Armandroff:1991aa}
{Armandroff} T.~E., {Da Costa} G.~S., 1991, \aj, 101, 1329

\bibitem[{{Bekki} \& {Freeman}(2003)}]{Bekki:2003aa}
{Bekki} K., {Freeman} K.~C., 2003, \mnras, 346, L11

\bibitem[{{Bekki} \& {Yong}(2012)}]{Bekki:2012aa}
{Bekki} K., {Yong} D., 2012, \mnras, 419, 2063

\bibitem[{{Bergemann} {et~al}\mbox{.}(2012){Bergemann}, {Lind}, {Collet},
  {Magic}, \& {Asplund}}]{Bergemann:2012aa}
{Bergemann} M., {Lind} K., {Collet} R., {Magic} Z., {Asplund} M., 2012, \mnras,
  427, 27

\bibitem[{{Carretta} {et~al}\mbox{.}(2009{\natexlab{a}}){Carretta},
  {Bragaglia}, {Gratton}, {D'Orazi}, \& {Lucatello}}]{Carretta:2009ac}
{Carretta} E., {Bragaglia} A., {Gratton} R., {D'Orazi} V., {Lucatello} S.,
  2009{\natexlab{a}}, \aap, 508, 695

\bibitem[{{Carretta} {et~al}\mbox{.}(2010{\natexlab{a}}){Carretta},
  {Bragaglia}, {Gratton}, {Lucatello}, {Bellazzini}, {Catanzaro}, {Leone},
  {Momany}, {Piotto}, \& {D'Orazi}}]{Carretta:2010ab}
{Carretta} E. {et~al.}, 2010{\natexlab{a}}, \aap, 520, A95

\bibitem[{{Carretta} {et~al}\mbox{.}(2010{\natexlab{b}}){Carretta},
  {Bragaglia}, {Gratton}, {Lucatello}, {Bellazzini}, {Catanzaro}, {Leone},
  {Momany}, {Piotto}, \& {D'Orazi}}]{Carretta:2010ac}
{Carretta} E. {et~al.}, 2010{\natexlab{b}}, \apjl, 714, L7

\bibitem[{{Carretta} {et~al}\mbox{.}(2009{\natexlab{b}}){Carretta},
  {Bragaglia}, {Gratton}, {Lucatello}, {Catanzaro}, {Leone}, {Bellazzini},
  {Claudi}, {D'Orazi}, {Momany}, {Ortolani}, {Pancino}, {Piotto},
  {Recio-Blanco}, \& {Sabbi}}]{Carretta:2009ab}
{Carretta} E. {et~al.}, 2009{\natexlab{b}}, \aap, 505, 117

\bibitem[{{Carretta} {et~al}\mbox{.}(2011){Carretta}, {Lucatello}, {Gratton},
  {Bragaglia}, \& {D'Orazi}}]{Carretta:2011aa}
{Carretta} E., {Lucatello} S., {Gratton} R.~G., {Bragaglia} A., {D'Orazi} V.,
  2011, \aap, 533, A69

\bibitem[{{Casetti-Dinescu} {et~al}\mbox{.}(2010){Casetti-Dinescu}, {Girard},
  {Korchagin}, {van Altena}, \& {L{\'o}pez}}]{Casetti-Dinescu:2010aa}
{Casetti-Dinescu} D.~I., {Girard} T.~M., {Korchagin} V.~I., {van Altena} W.~F.,
  {L{\'o}pez} C.~E., 2010, \aj, 140, 1282

\bibitem[{{Cohen}, {Huang} \& {Kirby}(2011){Cohen}, {Huang}, \&
  {Kirby}}]{Cohen:2011aa}
{Cohen} J.~G., {Huang} W., {Kirby} E.~N., 2011, \apj, 740, 60

\bibitem[{{Cohen} \& {Kirby}(2012)}]{Cohen:2012aa}
{Cohen} J.~G., {Kirby} E.~N., 2012, \apj, 760, 86

\bibitem[{{Cohen} {et~al}\mbox{.}(2010){Cohen}, {Kirby}, {Simon}, \&
  {Geha}}]{Cohen:2010aa}
{Cohen} J.~G., {Kirby} E.~N., {Simon} J.~D., {Geha} M., 2010, \apj, 725, 288

\bibitem[{{Cunha} {et~al}\mbox{.}(2002){Cunha}, {Smith}, {Suntzeff}, {Norris},
  {Da Costa}, \& {Plez}}]{Cunha:2002aa}
{Cunha} K., {Smith} V.~V., {Suntzeff} N.~B., {Norris} J.~E., {Da Costa} G.~S.,
  {Plez} B., 2002, \aj, 124, 379

\bibitem[{{Da Costa}(2015)}]{Da-Costa:2015aa}
{Da Costa} G.~S., 2015, "The General Assembly of Galaxy Halos: Structure,
Origin and Evolution", Proceedings of IAU Symposium 317, eds A. Bragaglia, M.
Arnaboldi, M. Rejkuba and D. Romano, (Cambridge: CUP), in press
(arXiv:1510.00873) 

\bibitem[{{Da Costa}, {Held} \& {Saviane}(2014){Da Costa}, {Held}, \&
  {Saviane}}]{Da-Costa:2014aa}
{Da Costa} G.~S., {Held} E.~V., {Saviane} I., 2014, \mnras, 438, 3507

\bibitem[{{Da Costa} {et~al}\mbox{.}(2009){Da Costa}, {Held}, {Saviane}, \&
  {Gullieuszik}}]{Da-Costa:2009aa}
{Da Costa} G.~S., {Held} E.~V., {Saviane} I., {Gullieuszik} M., 2009, \apj,
  705, 1481

\bibitem[{{D'Orazi} {et~al}\mbox{.}(2011){D'Orazi}, {Gratton}, {Pancino},
  {Bragaglia}, {Carretta}, {Lucatello}, \& {Sneden}}]{DOrazi:2011aa}
{D'Orazi} V., {Gratton} R.~G., {Pancino} E., {Bragaglia} A., {Carretta} E.,
  {Lucatello} S., {Sneden} C., 2011, \aap, 534, A29

\bibitem[{{Ferraro} {et~al}\mbox{.}(2009){Ferraro}, {Dalessandro},
  {Mucciarelli}, {Beccari}, {Rich}, {Origlia}, {Lanzoni}, {Rood}, {Valenti},
  {Bellazzini}, {Ransom}, \& {Cocozza}}]{Ferraro:2009aa}
{Ferraro} F.~R. {et~al.}, 2009, \nat, 462, 483

\bibitem[{{Freeman}(1993)}]{Freeman:1993aa}
{Freeman} K.~C., 1993, in Astronomical Society of the Pacific Conference
  Series, Vol.~48, The Globular Cluster-Galaxy Connection, {Smith} G.~H.,
  {Brodie} J.~P., eds., p. 608

\bibitem[{{Freeman} \& {Rodgers}(1975)}]{Freeman:1975aa}
{Freeman} K.~C., {Rodgers} A.~W., 1975, \apjl, 201, L71

\bibitem[{{Gratton} {et~al}\mbox{.}(2012){Gratton}, {Villanova}, {Lucatello},
  {Sollima}, {Geisler}, {Carretta}, {Cassisi}, \& {Bragaglia}}]{Gratton:2012ac}
{Gratton} R.~G., {Villanova} S., {Lucatello} S., {Sollima} A., {Geisler} D.,
  {Carretta} E., {Cassisi} S., {Bragaglia} A., 2012, \aap, 544, A12

\bibitem[{{Grillmair} {et~al}\mbox{.}(1995){Grillmair}, {Freeman}, {Irwin}, \&
  {Quinn}}]{Grillmair:1995aa}
{Grillmair} C.~J., {Freeman} K.~C., {Irwin} M., {Quinn} P.~J., 1995, \aj, 109,
  2553

\bibitem[{{Han} {et~al}\mbox{.}(2015){Han}, {Lim}, {Seo}, \&
  {Lee}}]{Han:2015aa}
{Han} S.-I., {Lim} D., {Seo} H., {Lee} Y.-W., 2015, \apjl, 813, L43

\bibitem[{{Harris}(1996)}]{Harris:1996aa}
{Harris} W.~E., 1996, \aj, 112, 1487

\bibitem[{{Hook} {et~al}\mbox{.}(2004){Hook}, {J{\o}rgensen},
  {Allington-Smith}, {Davies}, {Metcalfe}, {Murowinski}, \&
  {Crampton}}]{Hook:2004aa}
{Hook} I.~M., {J{\o}rgensen} I., {Allington-Smith} J.~R., {Davies} R.~L.,
  {Metcalfe} N., {Murowinski} R.~G., {Crampton} D., 2004, \pasp, 116, 425

\bibitem[{{Johnson} \& {Pilachowski}(2010)}]{Johnson:2010aa}
{Johnson} C.~I., {Pilachowski} C.~A., 2010, \apj, 722, 1373

\bibitem[{{Johnson} {et~al}\mbox{.}(2015){Johnson}, {Rich}, {Pilachowski},
  {Caldwell}, {Mateo}, {Bailey}, \& {Crane}}]{Johnson:2015aa}
{Johnson} C.~I., {Rich} R.~M., {Pilachowski} C.~A., {Caldwell} N., {Mateo} M.,
  {Bailey}, III J.~I., {Crane} J.~D., 2015, \aj, 150, 63

\bibitem[{{Kacharov}, {Koch} \& {McWilliam}(2013){Kacharov}, {Koch}, \&
  {McWilliam}}]{Kacharov:2013aa}
{Kacharov} N., {Koch} A., {McWilliam} A., 2013, \aap, 554, A81

\bibitem[{{Kunder} {et~al}\mbox{.}(2014){Kunder}, {Bono}, {Piffl}, {Steinmetz},
  {Grebel}, {Anguiano}, {Freeman}, {Kordopatis}, {Zwitter}, {Scholz}, {Gibson},
  {Bland-Hawthorn}, {Seabroke}, {Boeche}, {Siebert}, {Wyse}, {Bienaym{\'e}},
  {Navarro}, {Siviero}, {Minchev}, {Parker}, {Reid}, {Gilmore}, {Munari}, \&
  {Helmi}}]{Kunder:2014aa}
{Kunder} A. {et~al.}, 2014, \aap, 572, A30

\bibitem[{{Lee}, {Gim} \& {Casetti-Dinescu}(2007){Lee}, {Gim}, \&
  {Casetti-Dinescu}}]{Lee:2007aa}
{Lee} Y.-W., {Gim} H.~B., {Casetti-Dinescu} D.~I., 2007, \apjl, 661, L49

\bibitem[{{Lee} {et~al}\mbox{.}(1999){Lee}, {Joo}, {Sohn}, {Rey}, {Lee}, \&
  {Walker}}]{Lee:1999aa}
{Lee} Y.-W., {Joo} J.-M., {Sohn} Y.-J., {Rey} S.-C., {Lee} H.-C., {Walker}
  A.~R., 1999, \nat, 402, 55

\bibitem[{{Lind}, {Bergemann} \& {Asplund}(2012){Lind}, {Bergemann}, \&
  {Asplund}}]{Lind:2012aa}
{Lind} K., {Bergemann} M., {Asplund} M., 2012, \mnras, 427, 50

\bibitem[{{Marcolini} {et~al}\mbox{.}(2007){Marcolini}, {Sollima}, {D'Ercole},
  {Gibson}, \& {Ferraro}}]{Marcolini:2007aa}
{Marcolini} A., {Sollima} A., {D'Ercole} A., {Gibson} B.~K., {Ferraro} F.~R.,
  2007, \mnras, 382, 443

\bibitem[{{Marino} {et~al}\mbox{.}(2015){Marino}, {Milone}, {Karakas},
  {Casagrande}, {Yong}, {Shingles}, {Da Costa}, {Norris}, {Stetson}, {Lind},
  {Asplund}, {Collet}, {Jerjen}, {Sbordone}, {Aparicio}, \&
  {Cassisi}}]{Marino:2015aa}
{Marino} A.~F. {et~al.}, 2015, \mnras, 450, 815

\bibitem[{{Marino} {et~al}\mbox{.}(2009){Marino}, {Milone}, {Piotto},
  {Villanova}, {Bedin}, {Bellini}, \& {Renzini}}]{Marino:2009aa}
{Marino} A.~F., {Milone} A.~P., {Piotto} G., {Villanova} S., {Bedin} L.~R.,
  {Bellini} A., {Renzini} A., 2009, \aap, 505, 1099

\bibitem[{{Marino} {et~al}\mbox{.}(2011{\natexlab{a}}){Marino}, {Milone},
  {Piotto}, {Villanova}, {Gratton}, {D'Antona}, {Anderson}, {Bedin}, {Bellini},
  {Cassisi}, {Geisler}, {Renzini}, \& {Zoccali}}]{Marino:2011aa}
{Marino} A.~F. {et~al.}, 2011{\natexlab{a}}, \apj, 731, 64

\bibitem[{{Marino} {et~al}\mbox{.}(2014){Marino}, {Milone}, {Yong}, {Dotter},
  {Da Costa}, {Asplund}, {Jerjen}, {Mackey}, {Norris}, {Cassisi}, {Sbordone},
  {Stetson}, {Weiss}, {Aparicio}, {Bedin}, {Lind}, {Monelli}, {Piotto},
  {Angeloni}, \& {Buonanno}}]{Marino:2014aa}
{Marino} A.~F. {et~al.}, 2014, \mnras, 442, 3044

\bibitem[{{Marino} {et~al}\mbox{.}(2011{\natexlab{b}}){Marino}, {Sneden},
  {Kraft}, {Wallerstein}, {Norris}, {da Costa}, {Milone}, {Ivans}, {Gonzalez},
  {Fulbright}, {Hilker}, {Piotto}, {Zoccali}, \& {Stetson}}]{Marino:2011ab}
{Marino} A.~F. {et~al.}, 2011{\natexlab{b}}, \aap, 532, A8

\bibitem[{{Mauro} {et~al}\mbox{.}(2014){Mauro}, {Moni Bidin}, {Geisler},
  {Saviane}, {Da Costa}, {Gormaz-Matamala}, {Vasquez}, {Chen{\'e}}, {Cohen}, \&
  {Dias}}]{Mauro:2014aa}
{Mauro} F. {et~al.}, 2014, \aap, 563, A76

\bibitem[{{Mucciarelli} {et~al}\mbox{.}(2012){Mucciarelli}, {Bellazzini},
  {Ibata}, {Merle}, {Chapman}, {Dalessandro}, \&
  {Sollima}}]{Mucciarelli:2012aa}
{Mucciarelli} A., {Bellazzini} M., {Ibata} R., {Merle} T., {Chapman} S.~C.,
  {Dalessandro} E., {Sollima} A., 2012, \mnras, 426, 2889

\bibitem[{{Mucciarelli} {et~al}\mbox{.}(2015){Mucciarelli}, {Lapenna},
  {Massari}, {Pancino}, {Stetson}, {Ferraro}, {Lanzoni}, \&
  {Lardo}}]{Mucciarelli:2015aa}
{Mucciarelli} A., {Lapenna} E., {Massari} D., {Pancino} E., {Stetson} P.~B.,
  {Ferraro} F.~R., {Lanzoni} B., {Lardo} C., 2015, \apj, 809, 128

\bibitem[{{Navin}, {Martell} \& {Zucker}(2015){Navin}, {Martell}, \&
  {Zucker}}]{Navin:2015aa}
{Navin} C.~A., {Martell} S.~L., {Zucker} D.~B., 2015, \mnras, 453, 531

\bibitem[{{Norris}(1987)}]{Norris:1987aa}
{Norris} J., 1987, \apjl, 313, L65

\bibitem[{{Norris} \& {Freeman}(1983)}]{Norris:1983aa}
{Norris} J., {Freeman} K.~C., 1983, \apj, 266, 130

\bibitem[{{Norris} \& {Da Costa}(1995)}]{Norris:1995aa}
{Norris} J.~E., {Da Costa} G.~S., 1995, \apj, 447, 680

\bibitem[{{Norris}, {Freeman} \& {Mighell}(1996){Norris}, {Freeman}, \&
  {Mighell}}]{Norris:1996aa}
{Norris} J.~E., {Freeman} K.~C., {Mighell} K.~J., 1996, \apj, 462, 241

\bibitem[{{Olszewski} {et~al}\mbox{.}(2009){Olszewski}, {Saha}, {Knezek},
  {Subramaniam}, {de Boer}, \& {Seitzer}}]{Olszewski:2009aa}
{Olszewski} E.~W., {Saha} A., {Knezek} P., {Subramaniam} A., {de Boer} T.,
  {Seitzer} P., 2009, \aj, 138, 1570

\bibitem[{{Origlia} {et~al}\mbox{.}(2013){Origlia}, {Massari}, {Rich},
  {Mucciarelli}, {Ferraro}, {Dalessandro}, \& {Lanzoni}}]{Origlia:2013aa}
{Origlia} L., {Massari} D., {Rich} R.~M., {Mucciarelli} A., {Ferraro} F.~R.,
  {Dalessandro} E., {Lanzoni} B., 2013, \apjl, 779, L5

\bibitem[{{Origlia} {et~al}\mbox{.}(2011){Origlia}, {Rich}, {Ferraro},
  {Lanzoni}, {Bellazzini}, {Dalessandro}, {Mucciarelli}, {Valenti}, \&
  {Beccari}}]{Origlia:2011aa}
{Origlia} L. {et~al.}, 2011, \apjl, 726, L20

\bibitem[{{Pancino} {et~al}\mbox{.}(2011){Pancino}, {Mucciarelli}, {Sbordone},
  {Bellazzini}, {Pasquini}, {Monaco}, \& {Ferraro}}]{Pancino:2011aa}
{Pancino} E., {Mucciarelli} A., {Sbordone} L., {Bellazzini} M., {Pasquini} L.,
  {Monaco} L., {Ferraro} F.~R., 2011, \aap, 527, A18

\bibitem[{{Robin} {et~al}\mbox{.}(2003){Robin}, {Reyl{\'e}}, {Derri{\`e}re}, \&
  {Picaud}}]{Robin:2003aa}
{Robin} A.~C., {Reyl{\'e}} C., {Derri{\`e}re} S., {Picaud} S., 2003, \aap, 409,
  523

\bibitem[{{Roederer}, {Marino} \& {Sneden}(2011){Roederer}, {Marino}, \&
  {Sneden}}]{Roederer:2011aa}
{Roederer} I.~U., {Marino} A.~F., {Sneden} C., 2011, \apj, 742, 37

\bibitem[{{Roederer} {et~al}\mbox{.}(2016){Roederer}, {Mateo}, {Bailey},
  {Spencer}, {Crane}, \& {Shectman}}]{Roederer:2016aa}
{Roederer} I.~U., {Mateo} M., {Bailey} J.~I., {Spencer} M., {Crane} J.~D.,
  {Shectman} S.~A., 2016, \mnras, 455, 2417

\bibitem[{{Romano} {et~al}\mbox{.}(2007){Romano}, {Matteucci}, {Tosi},
  {Pancino}, {Bellazzini}, {Ferraro}, {Limongi}, \& {Sollima}}]{Romano:2007aa}
{Romano} D., {Matteucci} F., {Tosi} M., {Pancino} E., {Bellazzini} M.,
  {Ferraro} F.~R., {Limongi} M., {Sollima} A., 2007, \mnras, 376, 405

\bibitem[{{Skrutskie} {et~al}\mbox{.}(2006){Skrutskie}, {Cutri}, {Stiening},
  {Weinberg}, {Schneider}, {Carpenter}, {Beichman}, {Capps}, {Chester},
  {Elias}, {Huchra}, {Liebert}, {Lonsdale}, {Monet}, {Price}, {Seitzer},
  {Jarrett}, {Kirkpatrick}, {Gizis}, {Howard}, {Evans}, {Fowler}, {Fullmer},
  {Hurt}, {Light}, {Kopan}, {Marsh}, {McCallon}, {Tam}, {Van Dyk}, \&
  {Wheelock}}]{Skrutskie:2006aa}
{Skrutskie} M.~F. {et~al.}, 2006, \aj, 131, 1163

\bibitem[{{Smith} {et~al}\mbox{.}(2000){Smith}, {Suntzeff}, {Cunha}, {Gallino},
  {Busso}, {Lambert}, \& {Straniero}}]{Smith:2000aa}
{Smith} V.~V., {Suntzeff} N.~B., {Cunha} K., {Gallino} R., {Busso} M.,
  {Lambert} D.~L., {Straniero} O., 2000, \aj, 119, 1239

\bibitem[{{Stanford} {et~al}\mbox{.}(2007){Stanford}, {Da Costa}, {Norris}, \&
  {Cannon}}]{Stanford:2007aa}
{Stanford} L.~M., {Da Costa} G.~S., {Norris} J.~E., {Cannon} R.~D., 2007, \apj,
  667, 911

\bibitem[{{Valenti}, {Ferraro} \& {Origlia}(2007){Valenti}, {Ferraro}, \&
  {Origlia}}]{Valenti:2007aa}
{Valenti} E., {Ferraro} F.~R., {Origlia} L., 2007, \aj, 133, 1287

\bibitem[{{Villanova}, {Geisler} \& {Piotto}(2010){Villanova}, {Geisler}, \&
  {Piotto}}]{Villanova:2010aa}
{Villanova} S., {Geisler} D., {Piotto} G., 2010, \apjl, 722, L18

\bibitem[{{White} \& {Shawl}(1987)}]{White:1987aa}
{White} R.~E., {Shawl} S.~J., 1987, \apj, 317, 246

\bibitem[{{Yong} \& {Grundahl}(2008)}]{Yong:2008ab}
{Yong} D., {Grundahl} F., 2008, \apjl, 672, L29

\bibitem[{{Yong} {et~al}\mbox{.}(2014){Yong}, {Roederer}, {Grundahl}, {Da
  Costa}, {Karakas}, {Norris}, {Aoki}, {Fishlock}, {Marino}, {Milone}, \&
  {Shingles}}]{Yong:2014aa}
{Yong} D. {et~al.}, 2014, \mnras, 441, 3396

\end{thebibliography}
\end{document}